\documentclass[a4paper,fleqn,usenatbib,useAMS]{mnras}
\usepackage{graphicx}
\usepackage{epsfig}
\usepackage{rotate}
\usepackage{amsmath}
\usepackage{amssymb}
\usepackage{amsfonts}
\usepackage{enumerate}
\usepackage{afterpage}
\usepackage{xcolor}
\usepackage{widetext}
\usepackage{bm}
\usepackage{hyperref}
\usepackage[T1]{fontenc}
\usepackage{ae,aecompl}

\renewcommand{\mathbf}{\bm}
\newcommand{\n}{\bm {n}}
\renewcommand{\k}{\bm {k}}
\newcommand{\x}{\bm {x}}
\newcommand{\ka}{\bm {k}_1}
\newcommand{\kb}{\bm {k}_2}
\newcommand{\kc}{\bm {k}_3}
\newcommand{\ko}{\mathcal{K}}
\newcommand{\kt}{\mathcal{K}^{(2)}}

\newcommand{\ep}{\varepsilon}

\newcommand{\cH}{\mathcal{H}}

\def\be{\begin{equation}}
\def\ee{\end{equation}}
\def\bea{\begin{eqnarray}}
\def\eea{\end{eqnarray}}
\def\i{\mathrm{i\,}}

\title{The dipole of the galaxy bispectrum}

\author[Chris Clarkson, Eline M. de Weerd, Sheean Jolicoeur, Roy Maartens, Obinna Umeh] {Chris Clarkson$^{1,2,3}$, Eline M. de Weerd$^{1}$, Sheean Jolicoeur$^{2}$, Roy Maartens$^{2,3,4}$, \and Obinna Umeh$^{4}$ \\
$^{1}$ School of Physics \& Astronomy, Queen Mary University of London, London E1 4NS, UK \\
$^{2}$ Department of Physics \& Astronomy, University of the Western Cape, Cape Town 7535, South Africa \\
$^{3}$ Department of Mathematics \& Applied Mathematics, University of Cape Town, Cape Town 7701, South Africa\\
$^{4}$Institute of Cosmology \& Gravitation, University of Portsmouth, Portsmouth PO1 3FX, UK}

\date{\today}

\begin{document}

\maketitle
\begin{abstract}

The bispectrum will play an important role in future galaxy surveys. On large scales it is a key probe for measuring primordial non-Gaussianity which can help differentiate between different inflationary models and other theories of the early universe. On these scales a variety of relativistic effects come into play once the galaxy number-count fluctuation is projected onto our past lightcone. 
We show for the first time that the leading relativistic correction from these distortions in the galaxy bispectrum generates a significant dipole, mainly from relativistic redshift space distortions. The amplitude of the dipole can be more than 10\% of the monopole even on equality scales. Such a dipole  is absent in the Newtonian approximation to the redshift space bispectrum, so it offers a clear signature of relativistic effects on cosmological scales in large scale structure.

\end{abstract}
\begin{keywords} cosmology: theory. cosmology: large-scale structure of Universe.
\end{keywords}

\subsection*{Introduction}

The bispectrum provides an increasingly important probe of large-scale structure, complementing the information in the power spectrum and improving constraints on cosmological parameters. It has the potential to detect primordial non-Gaussianity, a key goal of large-scale galaxy surveys. The inclusion of redshift space distortions (RSD) in the bispectrum is essential for this purpose
\citep{Verde:1998zr,Scoccimarro:1999ed}. Though this adds complexity, this means that more information can potentially be extracted~\citep{Tellarini:2016sgp}.

The dominant RSD effect on galaxy number counts at first order is given by $\delta_g(\k) = (b_{1} + f\mu^{2})\delta(\k)$, where $\mu=\n\cdot\hat{\k}$, with $\n$ the line of sight direction, $f$ the growth rate, and $b_1$ is the linear bias (we omit the dependence on redshift here and below for convenience). The leading correction to this effect is a Doppler term~\citep{Kaiser:1987qv,McDonald:2009dh,Challinor:2011bk} (see also \citet{Raccanelli:2016avd,Hall:2016bmm,Abramo:2017xnp})
proportional to $\bm{v}\cdot\n$, where $\bm{v}$ is the peculiar velocity:\footnote{\citet{Challinor:2011bk} provides the relativistic correction to the coefficient of  $\bm{v}\cdot\n$ given in \citet{Kaiser:1987qv,McDonald:2009dh}.}
\be \label{dopf}
\delta_g(\x) = b_{1} \delta(\x)-{1\over\cH}\partial_r(\bm{v}\cdot\n) +{A}\,\bm{v}\cdot\n ~~\to~~
\ee
\be \delta_g(\k)=\Big(b_1+f\mu^2+\i {A}\,f \mu{\cH\over k}\Big)\delta(\k)\,,\label{dopf2}
%\\
% {A} &=& {b_{\rm e}+{3\over2}\Omega_m -3 + (2-5s) \Big(1-{1\over r\cH} \Big)}\,.
\ee
where  ${A} = {b_{\rm e}+{3}\Omega_m/2 -3 + (2-5s) (1-{1/ r\cH} )}$.
Here  $b_{\rm e}=\partial (a^3 \bar{n}_g)/\partial \ln a$ is the evolution of comoving galaxy number density, $s=-(2/5)\partial \ln \bar{n}_g/\partial \ln L$ is the magnification bias ($L$ is the threshold luminosity), $r$ is the comoving radial distance ($\partial_r=\bm n\cdot\bm\nabla$) and we have assumed a $\Lambda$CDM background
($\cH'/\cH^2=1-3\Omega_m/2$, where $\cH$ is the conformal Hubble rate, a prime is differentiation with respect to conformal time, $\Omega_m$ is the evolving density contrast). In the  Fourier space expression~\eqref{dopf2} we can read off the relative contribution of each term by how they scale with $k$: terms like $\cH/k$ are suppressed on small scales when $\cH/k\ll1$ but become important around and above the equality scale. 
%(We use the plane parallel approximation throughout, so for very large scales wide angle effects need to be included.)

Although the galaxy density contrast \eqref{dopf2} is complex, the power spectrum is real:
\be
\big\langle \delta_g(\k)\delta_g(-\k)\big\rangle
=\Big[ \big(b_{1} + f\mu^{2}\big)^2+\Big(A\,f \mu{\cH\over k}\Big)^2\Big] \big\langle \delta(\k)\delta(-\k)\big\rangle \,,\nonumber
\ee
since $\mu_{-\k}=-\mu_{\k}$ enforces a cancellation of the imaginary part, and the RSD contribution is separate from the Doppler term.
  However, if we consider the cross-power spectrum for {\em two} matter tracers, this cancellation breaks down
%  , {since the correlator $\langle \delta_g(\k)\delta_{\tilde g}(-\k)\rangle$ is no longer an even function of $\mu$.} Then 
and  there is an imaginary part in the cross-power~ \citep{McDonald:2009dh,Bonvin:2014owa}:
\bea
P_{g \tilde g}(k)&=&\Big\{\Big[ \big(b_{1} + f\mu^{2}\big)\big(\tilde b_{1} + f\mu^{2}\big)+A\tilde A f^2\mu^2 {\cH^2\over k^2 }\Big]\nonumber\\&&
+\i f\mu\Big[\big(\tilde b_{1} + f\mu^{2}\big)A-\big(b_{1} + f\mu^{2}\big)\tilde A\Big]{\cH\over k} \Big\}P(k) \,.\nonumber
\eea
While the Doppler contribution to $P_g$ is $O((\cH/k)^{2})$,  the Doppler contribution to $P_{g\tilde g}$ mixes with the density and RSD to give an additional less suppressed part, i.e. $O(\cH/k)$. The nonzero multipoles of $P_g$ are $\ell=0,2,4$, whereas  $P_{g \tilde g}$ has a nonzero dipole (as well as an octupole).  There are also further relativistic corrections to this dipole part of the cross power spectrum~\citep{DiDio:2018zmk}.

A natural question is: what about the galaxy bispectrum? In the standard `Newtonian' approximation, with only RSD, the galaxy bispectrum for a single tracer at fixed redshift has no dipole, and only has even multipoles~\citep{Scoccimarro:1999ed,Nan:2017oaq}. But
with a lightcone corrected galaxy density contrast, the 3-point correlator, even for a {\em single} tracer, will no longer be an even function of $\k_a\!\cdot\n \,(a=1,2,3)$. In order to compute the consequent contribution to the galaxy bispectrum, \eqref{dopf} is not sufficient: we need its second-order generalisation, $\delta_g \to \delta_g+\delta^{(2)}_g/2$.

\subsection*{Relativistic contributions to the galaxy bispectrum}

At second order, the Doppler correction in  \eqref{dopf} generalises to $A\, \bm{v}^{(2)}\!\cdot\n$, but there are also quadratic coupling terms. The couplings involve not only the Doppler effect but also radial gradients of the potential (`gravitational redshift'), volume distortion effects, and second-order corrections to the density contrast. Most of these contributions are small, but those that scale as $(\cH/k)\delta^2$ are not, even on equality scales. Except on super-equality scales we can often neglect any terms $O((\cH/k)^{2})$ and higher, which makes the calculation considerably simpler. 

The leading correction can be extracted from the general expressions that include all relativistic  corrections to the Newtonian approximation, as given in  \citet{Bertacca:2014hwa} (see also \citet{Bertacca:2014dra,Yoo:2014sfa,DiDio:2014lka,Jolicoeur:2017nyt,DiDio:2018zmk}):  
\bea
&&\delta^{(2)}_{g{\rm D}}=A\, \bm{v}^{(2)}\!\!\cdot\n+2{C}(\bm{v}\cdot\n)\delta +2{{E}\over\cH}(\bm{v}\cdot\n)\partial_r(\bm{v}\cdot\n)\label{dg2}
\\&&\nonumber
+2{b_1\over\cH}\phi\, \partial_r\delta
+{2\over\cH^2}\big[\bm{v}\cdot\n \,\partial_r^2\phi-\phi\, \partial_r^2 (\bm{v}\cdot\n) \big] -{2\over \cH}\partial_r (\bm{v}\cdot\bm{v}), 
\eea
where $\phi$ is the gravitational potential,   $C = b_1(A + f)+{b_1'/ \cH}+ 2(1-{1/ r\cH} ){\partial b_1/ \partial\ln L}$ and
$E = {4-2A-{3\over2}\Omega_m}$. (This is in agreement with the independent re-derivation of the leading correction  given in~\citet{DiDio:2018zmk}. We have corrected a typo in the last bracket of line 1 of Eq. (2.15): $-f_{\rm evo}\to -2f_{\rm evo}\equiv -2b_{\rm e}$. Note that our $\n$ is minus theirs, and they use the convention $\delta_g+\delta^{(2)}_g$.) 
All but one of the contributions to this leading term contain Doppler contributions, so we label these terms with a D subscript. In this sense they can be thought of as the relativistic correction to redshift space distortions, but their origin is considerably more subtle than in the Newtonian picture~\citep{Bertacca:2014dra,DiDio:2018zmk}. These relativistic corrections all arise as projections along the line of sight $\n$. It is this projection that is responsible for the dipole in the observed bispectrum.  Beyond these leading terms in~\eqref{dg2} there are a host of local coupled terms which appear on larger scales. 
We follow most work on the Fourier bispectrum and neglect the effect of lensing magnification. This is reasonable for correlations at the same redshift and when using very thin redshift bins allowed by spectroscopic surveys \citep{DiDio:2018unb}. We also use the standard plane-parallel approximation, which is reasonable on ultra-large scales. However, we note that wide-angle effects in the power spectrum can be of the same order of magnitude as the Doppler-type effects in certain circumstances~\citep{Tansella:2017rpi}, and these should be incorporated in a more complete treatment.

The galaxy bispectrum  is defined in Fourier space by
\bea 
B_{g}( \mathbf{k}_{1},  \mathbf{k}_{2},  \k_3) &=& { \mathcal{K}( \k_{1})\mathcal{K}({\k}_{2}) \mathcal{K}^{(2)}(  \mathbf{k}_{1},  \mathbf{k}_{2}, \k_3)}
P(k_{1})P(k_{2}) \nonumber\\&&+\text{2 cyclic permutations}\,.\label{b6}
\eea
 The first-order kernel $\mathcal{K}=\mathcal{K}_{\rm N}+\mathcal{K}_{\rm D}$ is given by the term in brackets in \eqref{dopf2}.
At second order, $\mathcal{K}^{(2)}=\mathcal{K}_{\rm N}^{(2)}+\mathcal{K}_{\rm D}^{(2)}$, where
 the Newtonian kernel is~\citep{Verde:1998zr}
\bea
%\kt(\ka,\kb,\kc) &=&\kt_{\rm N}(\ka,\kb,\kc)+ \kt_{\rm D}(\ka,\kb,\kc) \label{d1}\\
&&\mathcal{K}_{\rm{N}}^{(2)}%(\mathbf{k}_{1},  \mathbf{k}_{2},\k_3) 
= b_{2}+ b_{1}F_{2}%(\mathbf{k}_{1}, \mathbf{k}_{2},\kc) 
%\nonumber\\&&~~
 -{2\over7}(b_1-1)S_{2}
 %(\mathbf{k}_{1}, \mathbf{k}_{2},\kc)
%\nonumber\\&&~~
+ f\,G_{2}%(\mathbf{k}_{1}, \mathbf{k}_{2},\kc)
\mu_3^{2}
+ {\cal Z}_2%({\k_{1}},  {\k_{2}},\kc)
\,.   \label{kn2}  %\\
\eea
Here $F_2({\k_{1}},  {\k_{2}},\kc),G_2({\k_{1}},  {\k_{2}},\kc)$ are the second-order density and velocity kernels, and ${\cal Z}_2({\k_{1}},  {\k_{2}},\kc)$ is the second-order  RSD kernel. We use a local bias model \citep{Desjacques:2016bnm}, which includes tidal bias with kernel  $S_2({\k_{1}},  {\k_{2}},\kc)$. The kernels are given in \citet{Tellarini:2016sgp}.
\begin{widetext}
The Doppler correction to \eqref{kn2} in Fourier space follows from \eqref{dg2}~\citep{Jolicoeur:2017eyi}:
%\footnote{This is consistent with the full relativistic second-order kernel given in [Jolicoeur2017].}
\bea
\mathcal{K}^{(2)}_{\mathrm{D}}(\bm{k}_{1}, \bm{k}_{2}, \bm{k}_{3}) &= & {\i} {\cH}
\bigg[-{3\over 2}\left(\mu_{1}{k_{1}\over k_2^2} + \mu_{2}{k_{2}\over k_1^2}\right)\Omega_m b_1 +2\mu_{12} \left({\mu_{1}\over k_2} + {\mu_{2} \over k_1}\right)f^2  + \left({\mu_{1}\over k_1} + {\mu_{2} \over k_2}\right){C} f  
\nonumber \\
&&~~~~~ -{3\over 2} \left(\mu_{1}^3{k_{1}\over k_2^2} + \mu_{2}^3{k_{2}\over k_1^2}\right)\Omega_mf 
 + \mu_{1}\mu_{2}\left({\mu_{1}\over k_2} + {\mu_{2} \over k_1}\right){\Big({3\over2}\Omega_m-E f\Big) f} + {\mu_{3} \over k_{3}}G_{2}(\bm{k}_{1}, \bm{k}_{2},\kc){A} f\bigg] ,~~~~  \label{d2}
\eea
where $\mu_{ab}=\hat{\k}_a\cdot \hat{\k}_b$ and $\mu_a=\hat{\k}_a\cdot\bm n$. 
 The Newtonian kernel \eqref{kn2} scales as $(\cH/k)^{0}$, while the Doppler kernel \eqref{d2} scales as $(\cH/k)$. 
Using \eqref{kn2} and \eqref{d2} in \eqref{b6}, and dropping terms that scale as $(\cH/k)^{2}$ and $(\cH/k)^{3}$,
we find that
\bea 
B_{g{\rm N}}( \mathbf{k}_{1},  \mathbf{k}_{2},  \k_3) &=&  \ko_{\rm N}(\ka)\ko_{\rm N}(\kb)\kt_{\rm N}(\ka,\kb,\kc)\,P(k_{1})P(k_{2})  +\text{2 cyclic permutations} \,,\\
B_{g{\rm D}}( \mathbf{k}_{1},  \mathbf{k}_{2},  \k_3) &=& \Big\{
\ko_{\rm N}(\ka)\ko_{\rm N}(\kb)\kt_{\rm D}(\ka,\kb,\kc)
\nonumber\\ \label{bd}
&&~~~ 
+\Big[ \ko_{\rm N}(\ka)\ko_{\rm D}(\kb)+\ko_{\rm D}(\ka)\ko_{\rm N}(\kb) \Big]\kt_{\rm N}(\ka,\kb,\kc)
\Big\}
P(k_{1})P(k_{2}) +\text{2 c.p.}
\eea
Since \eqref{d2} scales as $\cH/k$ it is purely imaginary, as all these contributions have at least one $\k$ projected along the line of sight~-- i.e.,  they contain odd powers of $\mu_a$'s. This means that {\em the leading relativistic correction in the observed galaxy Fourier bispectrum of a single tracer is a purely imaginary addition to the Newtonian approximation.} On larger scales, terms $O((\cH/k)^{2})$ and higher appear in both the real and imaginary parts, with the kernels given in~\citet{Umeh:2016nuh,Jolicoeur:2017nyt,Jolicoeur:2017eyi,Jolicoeur:2018blf}. (We include these in our plots below.)

\subsection*{Extracting the dipole}

The bispectrum can be considered as a function of
$k_1,k_2,k_3,\mu_1, \mu_2,\mu_3$ and $\varphi$, which is the azimuthal angle giving the orientation of the triangle  relative to $\n$. In order to extract the dipole it is easiest to write $\mu_3=-(k_1\mu_1+k_2\mu_2)/k_3$, so that we can write $B_g = \sum_{i,j} \mathcal{B}_{ij}(\i\mu_1)^i(\i\mu_2)^j$, where $i,j=0\ldots6$ which factors out the angular dependence multiplying real coefficients $\mathcal{B}_{ij}$ with no angular dependence. Then,
 use the identity 
$%\be
\mu_2=\mu_1\cos\theta+\sqrt{1-\mu_1^2}\,\sin\theta\,\cos\varphi\,,
$ %\ee 
where $\theta=\theta_{12}$ (and we define $\mu=\cos\theta$~-- note that $\theta$ is the angle outside the triangle as the $\k_a$'s are head-to-tail). We use standard orthonormal spherical harmonics with the triangle lying in the $y-z$ plane, with $\bm k_1$ aligned along the $z$-axis~\citep{Nan:2017oaq}. Then we have $Y_{\ell m}(\mu_1,\varphi)$, so that we can write $B_g =\sum_{\ell m} B_{\ell m}Y_{\ell m}(\mu_1,\varphi)$. The leading relativistic terms we consider here generate odd-power multipoles up to $\ell=7$, and the full expression generates even and odd multipoles up to $\ell=8$.
Different powers of $(\i\mu_1)$ and $(\i\mu_2)$ contribute to the dipole, 
\be
\arraycolsep=1.0pt\def\arraystretch{1}
\int\mathrm{d}\Omega (\i\mu_1)^i  (\i \mu_2)^j Y^*_{1 m} = 
 \delta_{m,0}\frac{\i\sqrt{3\pi}}{15}\left[ \begin {array}{ccccc} 
 0&10\mu&0&-6\mu  &\\ 
10&0&-4{\mu}^{2}-2&0& \cdots\\ 
0&-6\,\mu&0&{\frac {12{
\mu}^{3}+18\mu}{7}}& \\ 
-6&0&{\frac {24
\,{\mu}^{2}+6}{7}}&0& \\
 &\vdots & & & \ddots
\end {array} \right] 
+
\delta_{m,\pm1}\frac{\sqrt{6\pi}}{15}
 \left[ \begin {array}{ccccc} 0&-5&0&3&\\ 
 0&0&2\,\mu&0
& \cdots\\ 
0&1&0&-\frac{6{\mu}^{2}+3}{7}&\\ 
0&0
&-\frac{6}{7}\,\mu&0&\\
 &\vdots & & & \ddots
\end {array} \right] \sin\theta\,,
\label{dkjsncdjcnsk}
\ee
where each matrix element corresponds to a particular combination of $i,j$,
where the matrix indices run over the values $i=0\ldots6, j=0\ldots6$, with powers above 3 not written above; these are polynomials in $\mu$ up to order 6. From this we can read off the terms from $\mathcal{K}_\text{D}$ contribute to differing $m=0,\pm1$. In particular, if $i+j$ is even~-- i.e., the real part of the bispectrum~--  there is no contribution: only the imaginary terms, corresponding to $i+j$ odd, contribute. For the monopole, only $i+j$ even contribute. Therefore, at $O(\cH/k)$, \emph{the monopole of the bispectrum is the Newtonian part, while the dipole is purely from the relativistic corrections.  The presence of the dipole is therefore a `smoking gun' signal for the leading relativistic correction to the bispectrum.} At order $O((\cH/k)^{2})$, relativistic terms appear in the monopole, which were considered in \citet{Umeh:2016nuh,Jolicoeur:2017nyt,Jolicoeur:2017eyi,Jolicoeur:2018blf}.\\

\end{widetext}

\subsection*{Squeezed, equilateral and flattened limits}

It is relatively straightforward to understand the type of dipole generated in different triangular configurations in our conventions. In particular, for the $O(\cH/k)$ relativistic dipole:
\begin{itemize}
\item The squeezed case is zero for $m=0$, and is non-zero for $m=\pm1$. We see this directly from \eqref{dkjsncdjcnsk}: with $\mu=-1$ the $m=0$ contribution is anti-symmetric in $i,j$ while $\mathcal{B}_{ij}$ is symmetric in this limit.
\item In the equilateral case, the dipole is zero (this is the case for all orders in $\cH/k$).
\item The flattened case ($k_1=k_2=k_3/2,\theta=0$) is zero for $m=\pm1$ (for all orders in $\cH/k$), but is non-zero for $m=0$. This can be seen directly from \eqref{dkjsncdjcnsk} with $\theta=0$.
\end{itemize}
%Geometrically, we can understand these by rotating the triangle about its centre with respect to $\bm n$. For the equilateral case, the symmetry of the triangle means that the dipole part cancels out. For the flattened case, where there is a vertex at the centre, an anti-symmetry occurs only parallel to $\bm n$, and cancels in the plane orthogonal. For the squeezed case, this is anti-symmetric both along and perpendicular to $\bm n$, so excites both $m=\pm1$. 
 
To show the equilateral case is zero is a lengthy calculation involving many cancellations.  Let us illustrate instead the squeezed case. We write $k_1=k_2=\sqrt{1+\varepsilon^2}k_S, k_3=2\ep k_S$.
In this case the triangle has small angle $2\ep$ and equal angles $\pi/2-\ep$, where the squeezed limit is $\ep\to0$. It is convenient to replace $(1,2,3)$ by $(S,-S,L)$.
Then to $O(\ep)$,
$%\bea
k_{-S}= k_{S}\,,~~k_L=2\ep k_S\,,~~  
\mu_{-S}=-\mu_S-2\ep\mu_L\,,
%\nonumber\\&&
\mu_L = -\sqrt{1-\mu_S^2}\,\cos\varphi - {\ep\mu_S}\,.
%\hat{\k}_2=-\hat{\k}_1-2\ep\,\hat{\k}_3 \\ &&
$% \label{mu}
%\eea
 ~In this limit, the permutations of the relativistic kernels become
\bea
% {\cal K}^{(2)}_{\rm D}(\k_S,\k_{-S},\k_L) &=&0\\
&& {\cal K}^{(2)}_{\rm D}(\k_L,\k_S,\k_{-S}) = \i {\cH}
{\bigg[-{3\over2}\Omega_m b_1\mu_S{k_S \over  k_L^2}+ C f{\mu_L\over k_L} }
\nonumber\\&&
-{3\over2}\Omega_m f\mu_S^3{k_S\over k_L^2} +\Big({3\over2}\Omega_m- Ef \Big) f\mu_S^2{\mu_L\over k_L}\bigg] 
\label{k2d1} 
% {\cal K}^{(2)}_{\rm D}(\k_{-S},\k_L,\k_S) &=&  {\cal K}^{(2)}_{\rm D}(\k_L,\k_S,\k_{-S})\Big|_{\mu_S\to \roy{\mu_{-S}}}  \label{k2d2}
\eea
and $ {\cal K}^{(2)}_{\rm D}(\k_{-S},\k_L,\k_S) =  {\cal K}^{(2)}_{\rm D}(\k_L,\k_S,\k_{-S})\big|_{\mu_S\to {\mu_{-S}}}$ while ${\cal K}^{(2)}_{\rm D}(\k_S,\k_{-S},\k_L) =0$. 
In the squeezed limit of the cyclic sum  \eqref{b6}, the terms $ {\cal K}^{(2)}(\k_L,\k_S,\k_{-S})$ and  ${\cal K}^{(2)}(\k_{-S},\k_L,\k_S)$ appear only in the form ${\cal K}^{(2)}(\k_L,\k_S,\k_{-S})+{\cal K}^{(2)}(\k_{-S},\k_L,\k_S)$. This sum regularises
the divergent $k_S/k_L=(2\ep)^{-1}$ and $k_S/k_L^2=(2\ep k_L)^{-1}$ terms.  We obtain the bispectrum in the squeezed limit,
\bea
%B_{g}^{\rm sq}
%&=& B_{g {\rm N}}^{\rm sq}+B_{g {\rm D}}^{\rm sq} \nonumber\\
%B_{g {\rm N}}^{\rm sq} &=& b_{1S}^2\Big(b_2-{4\over21}b_1+{4\over 21} \Big)P_S^2+b_{1S}b_{1L}b_{SL}\, P_LP_S\,,\nonumber
&&B_{g {\rm}}^{\rm sq} = b_{1S}b_{1L}b_{SL}\, P_LP_S+\i b_{1S}\Big\{b_{SL}f A + 
%\big[-\big( \beta_{10}+3\mu_S^2 \beta_{13}\big)+ 2\big( \beta_{12}+\mu_S^2 \beta_{14}\big)\big] 
{3\over2}\Omega_m b_{1S}b_{1L} 
\nonumber\\&&+2b_{1L}f C+ b_{1L}\mu_S^2\Big[{3\over2}\Omega_m- Ef \Big]}
\Big\}P_LP_S\,\mu_L{{\cH}\over k_L \,,\label{bgd}
 \eea
where $P_{S,L}=P(k_{S,L})$, $b_{1S,L}\equiv  b_1+f\mu_{S,L}^2$ and
\bea
% \,, \label{b1sl}~~
b_{SL} \equiv 2b_2+ {43\over 21}b_1-{4\over21}
%\nonumber\\&&
+\Big(2b_1+ {5\over7}\Big)f\mu_S^2%+b_1f\mu_L^2+{f^2}\mu_S^2\mu_L^2 
+f\mu_L^2 b_{1S}
\,. \nonumber\label{bsl}
\eea
%Note that we can neglect the $P(k_S)^2$ term relative to the $P(k_L)P(k_S)$ term in $B_{g {\rm N}}^{\rm sq}$.
Note that only the first term in the squeezed bispectrum comes from the Newtonian limit. 

The type of dipole extracted from this term is seen as follows. To this order we can write $\mu_{S}^2=\mu_{S}\mu_{-S}$. Then,  since $\mu_L=-2(\mu_S+\mu_{-S})/
\varepsilon$, we see that the $m=0$ term is zero because $B_{g {\rm D}}^{\rm sq}$ is symmetric in $\mu_{S}^i\mu_{-S}^j$ under $i\leftrightarrow j$, while the $m=0$ term is antisymmetric in~\eqref{dkjsncdjcnsk}. This leaves just the $m=\pm1$ contribution in \eqref{dkjsncdjcnsk}.

\begin{figure}%[htbp]
\begin{center}
\includegraphics[width=\columnwidth]{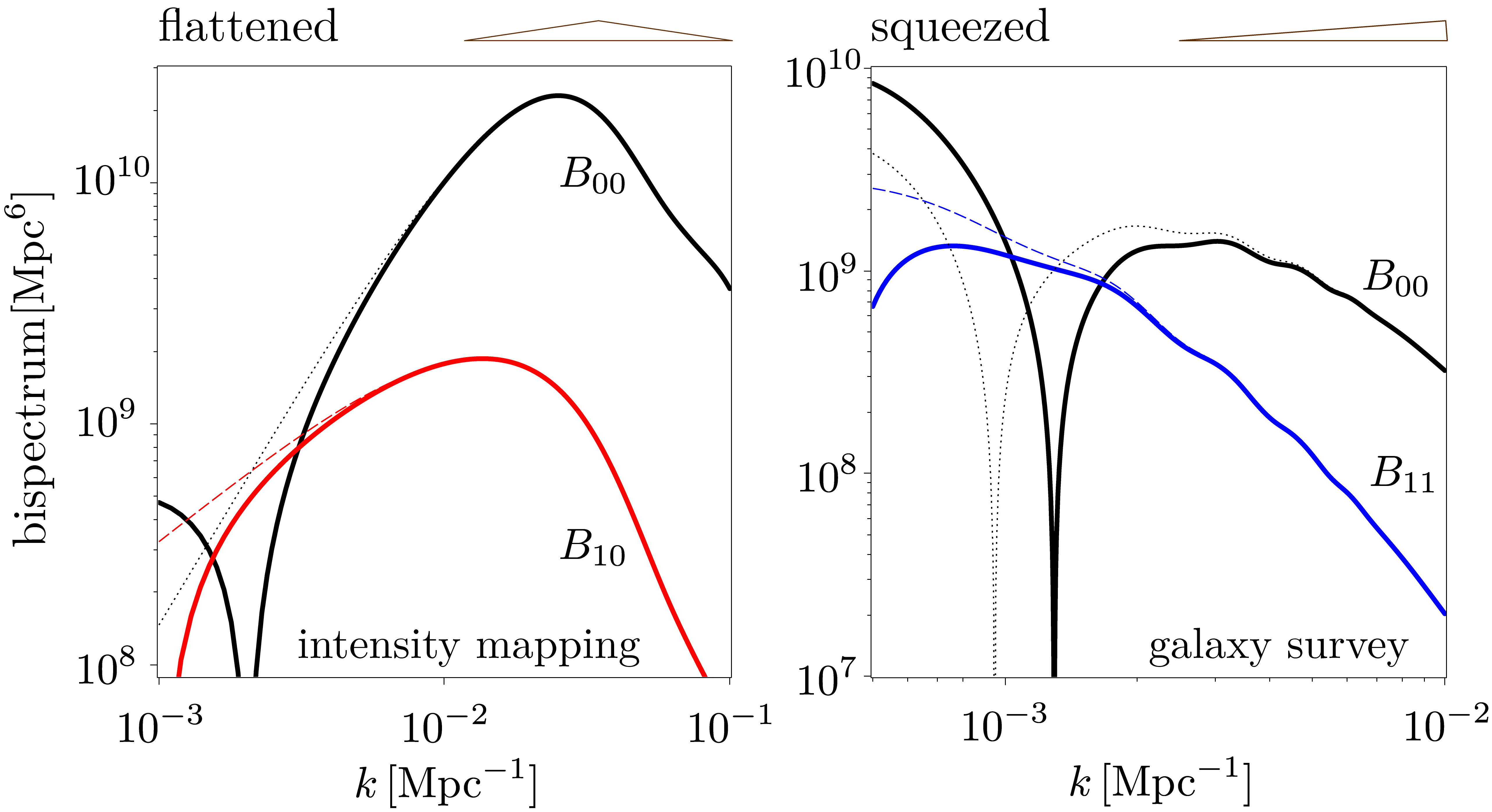}
\caption{ The absolute value of the bispectrum dipole at $z=1$  as a function of triangle size, in the flattened (Left, $\theta=2^\circ$, for intensity mapping bias) and squeezed (Right, $\theta=178^\circ$, for Euclid-like bias) configurations, with $k_3$ as the horizontal axis. Red is the $m=0$ part and blue is $m=\pm1$. Dashed (and dotted) lines show up to the $O(\cH/k)$ terms considered analytically here, while solid lines indicate larger-scale contributions. For reference the monopole is in black, with the dotted line the Newtonian part.  (The zero-crossing in the monopole for the squeezed case is a result of the tidal bias.)}
\label{snakcjnsdlkcans}
\end{center}
\end{figure}

\begin{figure}%[htbp]
\begin{center}
\includegraphics[width=\columnwidth]{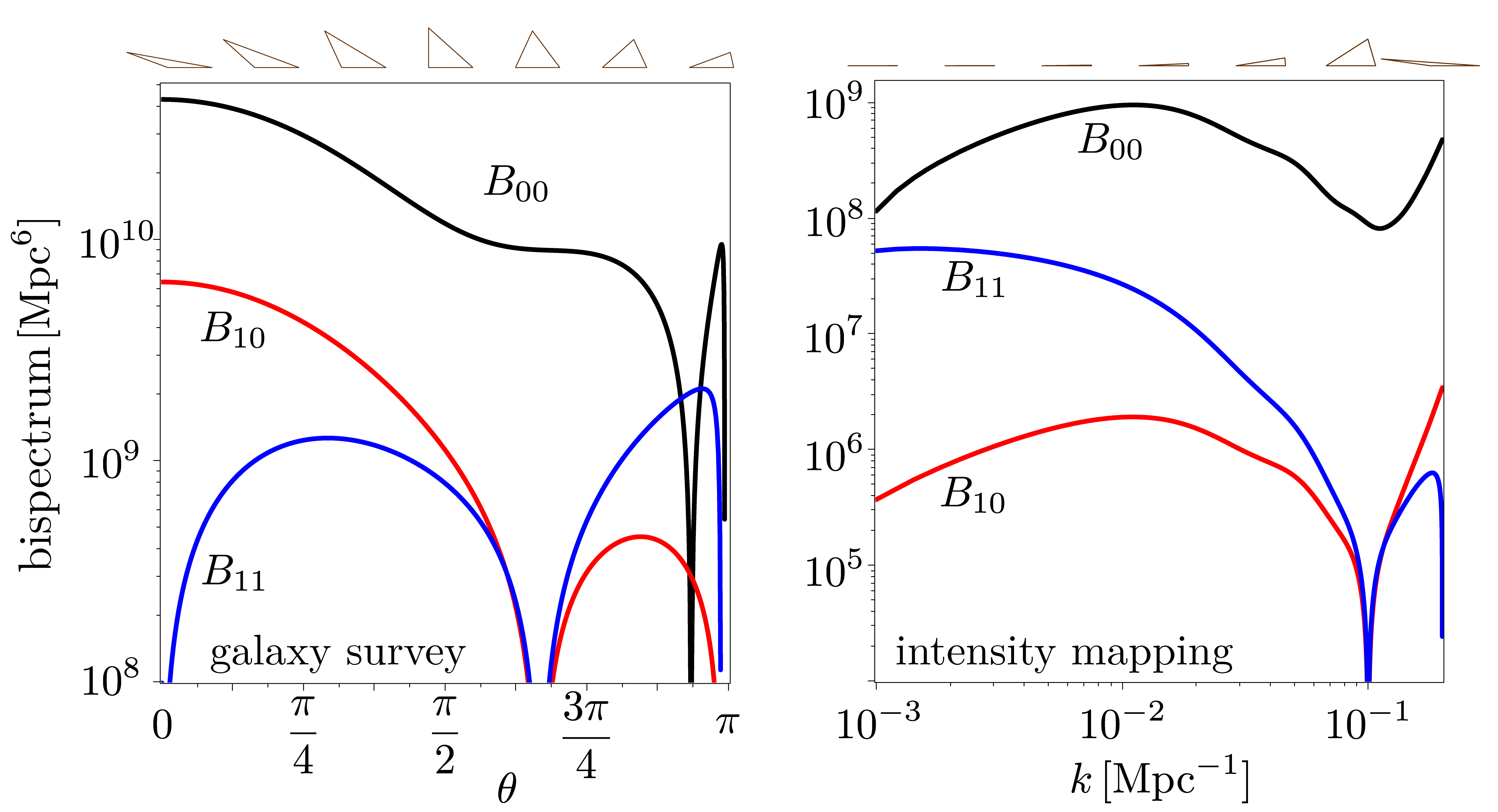}
\caption{ (Left) We show the dipoles as a function of $\theta$ with a bias appropriate for a Euclid-like survey, for $k_1=k_2=0.01$\,Mpc$^{-1}$. The left of the plot corresponds to the flattened case where the $m=0$ (red) dipole reaches 10\% of the monopole.  (Right) We show the IM signal with $k_1=k_2=0.1$\,Mpc$^{-1}$ versus the long mode $k_3$. Except for very long modes $\theta\approx\pi$, our $O(\cH/k)$ truncation is a very good approximation in these examples. }
\label{sankcjnakjdcs}
\end{center}
\end{figure}

\subsection*{The dipole in intensity mapping and galaxy surveys}

We now consider the amplitude of the dipole relevant for upcoming galaxy surveys, which have different bias parameters. We consider two different types of survey: an SKA intensity mapping of 21\,cm radio emission, as well as a Euclid-like optical/infrared spectroscopic survey.
An intensity map of the 21cm emission of neutral hydrogen (HI) in the post-reionization Universe records the total emission in galaxies containing HI, without detecting individual galaxies. There is an equivalence between the brightness temperature contrast and number count contrast~\citep{Umeh:2015gza}. For IM we use the bias parameters at $z=1$, 
$b_1 = 0.856, b_2 = -0.321, b_1' = -0.5\times10^{-4}, b_e = -0.5, b_e'=0, s = 2/5$~\citep{Fonseca:2018hsu,Umeh:2015gza}
while for the spectroscopic survey we use 
$ b_1 = 1.3,b_2 = -0.74, b_1' = -1.6\times10^{-4},  b_e = -4, b_e' = 0, s = -0.95$~\citep{Camera:2018jys,Yankelevich:2018uaz}.
For intensity mapping, $ \partial b_1/\partial ln L =0$ and we assume it is zero for simplicity for the spectroscopic survey. We use a LCDM model with standard parameters $\Omega_m=0.314, h=0.67, f_\text{baryon}=0.157, n_s=0.968$. Plots are presented using linear power spectra generated using CAMB~\citep{Lewis:1999bs}.

In Fig.~\ref{snakcjnsdlkcans} we show how changing the scale of a fixed triangle changes the amplitude of the dipole, with reference to the monopole. In the flattened case with $m=0$ we see the signal peaks for triangles below the equality scale, while for squeezed shapes, with $m=\pm1$, the signal is smaller, and peaks when the long mode approaches the Hubble scale. 
In Fig.~\ref{sankcjnakjdcs} we change the shape with fixed $k_1=k_2$ for both galaxy and IM surveys. We confirm our analytical results that the equilateral limit is zero, as well as the other limits. For triangles between right-angle and flattened the dipole is more than 10\% of the monopole, and the signal is largest in the flattened case~-- except in the extreme squeezed limit (not shown).

\subsection*{Conclusions}

We have shown for the first time that the relativistic galaxy bispectrum has a leading correction which is a local dipole with respect to the observers line of sight. In contrast to the power spectrum, this dipole exists even for a single tracer. We have shown analytically how the dipole is generated for the leading terms, and numerically we have included all local contributions, which show up above the equality scale. We have neglected integrated terms which will also contribute to the dipole, but their inclusion in a Fourier space bispectrum is non-trivial. Local relativistic corrections will induce all multipoles up to $\ell=8$ at every $m$, in contrast to the Newtonian case which only induces even $\ell=0,2,4$. We will investigate these new multipoles in a forthcoming publication. 

We have shown that this dipole is large with respect to the monopole in both the flattened and squeezed limits, which excite different orders of the dipole orientation $m$.  We have shown that even on equality scales it is about 10\% of the monopole at $z=1$ for flattened shapes which have the largest amplitude. In more squeezed cases where the short mode is $\sim10$\,Mpc the dipole can also be a large part of the IM signal. Furthermore, although we have only considered Gaussian initial conditions here, the dipole will be unaffected by non-Gaussianity at leading order because these corrections start at $O((\cH/k)^2)$, making our predictions relatively robust to this. This implies that the dipole of the bispectrum is a unique signature of general relativity on cosmological scales, and therefore offers a new observational window onto modifications of general relativity.

~\\

\section*{Acknowledgements} CC was supported by STFC Consolidated Grant ST/P000592/1.  RM is supported by STFC Consolidated Grant ST/N000668/1, and by  the South African Radio Astronomy Observatory (SARAO) and the National Research Foundation (Grant No. 75415). OU is supported by STFC grant ST/N000668/1.

\bibliographystyle{mnras}
\bibliography{ref}

\end{document}